\documentclass[12pt,a4paper]{article}
\usepackage{amssymb,amsmath}
\allowdisplaybreaks
\begin{document}
\title{Finslerian 3-spinors and the generalized Duffin--Kemmer equation}
\author{A. V. Solov'yov}
\date{\normalsize\textit{Division of Theoretical Physics, Faculty of
  Physics,\\ Moscow State University, Moscow, Russia}}
\maketitle
\begin{abstract}
The main facts of the geometry of Finslerian 3-spinors are
formulated. The close connection between Finslerian 3-spinors and
vectors of the 9-dimensional linear Finslerian space is established. The
isometry group of this space is described. The procedure of
dimensional reduction to 4-dimensional quantities is formulated. The
generalized Duffin--Kemmer equation for a Finslerian 3-spinor wave
function of a free particle in the momentum representation is
obtained. From the viewpoint of a 4-dimensional observer, this
9-dimensional equation splits into the standard Dirac and
Klein--Gordon equations.
\end{abstract}
\section*{Introduction}
In the works \cite{Finkelstein:Hyperspin, Finkelstein:Hypermanifolds},
\textit{hyperspinors} and their basic properties were considered. The
same mathematical objects were independently studied under the name of
$N$-component spinors in the papers \cite{Vladimirov:3-spinors,
Solov'yov:4-spinors}. Finally, in the work \cite{Solo_Vlad:N-spinors},
the general algebraic theory of \textit{Finslerian $N$-spinors} was
constructed. The last term is more suitable because it reflects the
close connection between hyperspinors and Finslerian geometry.

This paper is devoted to formulating the main facts of the geometry of
Finslerian 3-spinors and deducing the generalized Duffin--Kemmer
equation for a Finslerian 3-spinor wave function of a free particle in
the momentum representation.

In short, the paper has the following structure. We begin with the
definition of the space of Finslerian 3-spinors and construct the
associated Finslerian geometry. After deducing the expression for the
length of a vector in the 9-dimensional linear Finslerian space, we
describe the corresponding isometry group. We also formulate the
procedure of dimensional reduction which allows us to rewrite the
expression for the Finslerian length of a 9-vector in terms of
4-dimensional geometric objects. We end the paper with deducing the
generalized Duffin--Kemmer equation for a Finslerian 3-spinor wave
function of a free particle in the momentum representation.
\section*{The geometry of Finslerian 3-spinors}
Let $\mathbb{C}^3$ be the linear space of 3-component columns of complex numbers
with respect to the standard matrix operations of addition and multiplication by
elements of the field $\mathbb{C}$. Let us consider the antisymmetric 3-linear
form
\begin{equation}
[\xi,\eta,\lambda]=\varepsilon_{abc}\,\xi^a\eta^b\lambda^c,
\label{eq:1}
\end{equation}
where $\xi$, $\eta$, $\lambda\in\mathbb{C}^3$, $\varepsilon_{abc}$ is
the Levi-Civita symbol with the ordinary normalization $\varepsilon_{123}=1$,
the indices $a$, $b$, $c$ run independently from 1 to 3, and $\xi^a$,
$\eta^b$, $\lambda^c\in\mathbb{C}$. Here and in the following formulas,
the summation is taken over all the repeating indices.

The space $\mathbb{C}^3$ equipped with the form \eqref{eq:1} is called the
\textit{space of Finslerian 3-spinors}. The complex number
$[\xi,\eta,\lambda]$ is respectively called the \textit{symplectic
scalar 3-product} of the Finslerian 3-spinors $\xi$, $\eta$, and
$\lambda$.

Since \eqref{eq:1} is the determinant
\begin{equation}
[\xi,\eta,\lambda]=
\begin{vmatrix}
\xi^1&\eta^1&\lambda^1\\
\xi^2&\eta^2&\lambda^2\\
\xi^3&\eta^3&\lambda^3
\end{vmatrix}
\label{eq:2}
\end{equation}
with the columns $\xi$, $\eta$, and $\lambda$, the symplectic scalar
3-product $[\xi,\eta,\lambda]$ vanishes if and only if the Finslerian
3-spinors $\xi$, $\eta$, and $\lambda$ are linearly dependent
\cite{Kostrikin}. In particular, $[\xi,\xi,\xi]=0$ for any
$\xi\in\mathbb{C}^3$.

Let us find isometries of the space of Finslerian 3-spinors, i.e., the linear
transformations
\begin{equation}
\xi^\prime=D\xi\quad\Longleftrightarrow\quad\xi^{\prime a}=d^a_b\xi^b\quad
(D=\|d^a_b\|; d^a_b\in\mathbb{C}; a,b=1,2,3)
\label{eq:3}
\end{equation}
which preserve the symplectic scalar 3-product:
\begin{equation}
[\xi^\prime,\eta^\prime,\lambda^\prime]=[\xi,\eta,\lambda]
\quad\text{for any}\quad\xi,\eta,\lambda\in\mathbb{C}^3.
\label{eq:4}
\end{equation}
Substituting \eqref{eq:3} and the similar expressions for $\eta^\prime$,
$\lambda^\prime$ into the condition \eqref{eq:4}, we obtain
\begin{equation}
[\xi,\eta,\lambda]\det D=[\xi,\eta,\lambda]
\label{eq:5}
\end{equation}
with regard to \eqref{eq:2}. Due to arbitrariness of $\xi$, $\eta$,
$\lambda\in\mathbb{C}^3$, the equation \eqref{eq:5} implies $\det
D=1$. Thus, the isometries of the space of Finslerian 3-spinors form
the group $\text{SL}(3,\mathbb{C})$.

Let us consider the subspace of the linear space $\mathbb{C}^3\otimes
\overline{\mathbb{C}^3}$ which consists of Hermitian tensors. This subspace is
isomorphic to the 9-dimensional \textit{real} linear space $\text{Herm}(3)=
\{X\mid X=X^+\}$ of all Hermitian $3\times 3$ matrices with complex
elements. Here and below, the over-line denotes complex conjugating,
while the cross does Hermitian conjugating.

As a basis of the space $\text{Herm}(3)$, we choose the following linearly
independent matrices
\begin{align}
\lambda_0&=
\begin{pmatrix}
1&0&0\\
0&1&0\\
0&0&0
\end{pmatrix},&
\lambda_1&=
\begin{pmatrix}
0&1&0\\
1&0&0\\
0&0&0
\end{pmatrix},&
\lambda_2&=
\begin{pmatrix}
0&-i&0\\
i&0&0\\
0&0&0
\end{pmatrix},\notag\\
\lambda_3&=
\begin{pmatrix}
1&0&0\\
0&-1&0\\
0&0&0
\end{pmatrix},&
\lambda_4&=
\begin{pmatrix}
0&0&1\\
0&0&0\\
1&0&0
\end{pmatrix},&
\lambda_5&=
\begin{pmatrix}
0&0&-i\\
0&0&0\\
i&0&0
\end{pmatrix},\notag\\
\lambda_6&=
\begin{pmatrix}
0&0&0\\
0&0&1\\
0&1&0
\end{pmatrix},&
\lambda_7&=
\begin{pmatrix}
0&0&0\\
0&0&-i\\
0&i&0
\end{pmatrix},&
\lambda_8&=
\begin{pmatrix}
0&0&0\\
0&0&0\\
0&0&1
\end{pmatrix}
\label{eq:7}
\end{align}
($\lambda_1,\lambda_2,\dots,\lambda_7$ are the well-known Gell-Mann
matrices). Then, for any $X\in\text{Herm}(3)$, we have the expansion
\begin{equation}
X=X^A\lambda_A\quad (A=0,1,\dots,8),
\label{eq:8}
\end{equation}
where $X^A\in\mathbb{R}$ are components of the 9-vector $X$ with respect to the
basis \eqref{eq:7}. Along with the matrices \eqref{eq:7}, we introduce another
set of the Hermitian $3\times 3$ matrices: $\lambda^B=\lambda_B$ ($B\ne 8$),
$\lambda^8=2\lambda_8$. Under such a choice of the matrices, the
remarkable relations
\begin{equation}
\text{Tr}(\lambda^A\lambda_B)=2\delta^A_B\quad (A,B=0,1,\dots,8)
\label{eq:9}
\end{equation}
are fulfilled. Here, $\text{Tr}(\lambda^A\lambda_B)$ denotes the trace
of the matrix $\lambda^A\lambda_B$ and $\delta^A_B$ is the Kronecker
symbol. Because of \eqref{eq:8} and \eqref{eq:9},
\begin{equation}
X^A=\frac{1}{2}\text{Tr}(\lambda^A X).
\label{eq:10}
\end{equation}

Let us equip $\text{Herm}(3)$ with the structure of the Finslerian space. To
this end, we define the \textit{length} $|X|$ of the 9-vector
$X\in\text{Herm}(3)$ in the following way:
$$
|X|\equiv\sqrt[3]{\det X}.
$$
Computing the determinant of \eqref{eq:8}, we obtain the expression for
$|X|^3$ in the basis \eqref{eq:7}:
\begin{align}
|X|^3&=G_{ABC}X^A X^B X^C=[(X^0)^2-(X^1)^2-(X^2)^2-(X^3)^2]X^8-\notag\\
&-X^0[(X^4)^2+(X^5)^2+(X^6)^2+(X^7)^2]+2X^1[X^4X^6+X^5X^7]+\notag\\
&+2X^2[X^5X^6-X^4X^7]+X^3[(X^4)^2+(X^5)^2-(X^6)^2-(X^7)^2].
\label{eq:11}
\end{align}
Here, $G_{ABC}$ are components of the covariant symmetric tensor on
$\text{Herm}(3)$. Thus, the Finslerian length of the 9-vector
$X\in\text{Herm}(3)$ in the basis \eqref{eq:7} is the form of degree 3 with
respect to its components \eqref{eq:10}. It should be noted that the form
\eqref{eq:11} is indefinite, i.e., the cases $|X|^3>0$, $|X|^3<0$, or
$|X|^3=0$ are possible. Since $|X|^3=\det X$, we have $|X|^3=0$ if and only if
$\det X=0$.

Any linear transformation \eqref{eq:3} of the space of Finslerian 3-spinors
induces the transformation
\begin{equation}
X^\prime=DXD^+\quad\Longleftrightarrow\quad
X^{\prime a\dot b}=d^a_c\overline{d^{\dot b}_{\dot e}}X^{c\dot e}\quad
(X^\prime=\|X^{\prime a\dot b}\|; X=\|X^{c\dot e}\|)
\label{eq:12}
\end{equation}
in $\text{Herm}(3)$. Here, all the indices run from 1 to 3 and
$X\in\text{Herm}(3)$. It is evident that the transformation \eqref{eq:12} has
the following properties:
\begin{enumerate}
\item If $X=X^+$, then $X^\prime=X^{\prime+}$, i.e.,
  $X\in\text{Herm}(3)$ implies $X^\prime\in\text{Herm}(3)$.
\item The transformation \eqref{eq:12} is linear with respect to $X$.
\item If $\det D=1$, then $\det X^\prime=\det X$ for any $X\in\text{Herm}(3)$.
\end{enumerate}
Since $|X|=\sqrt[3]{\det X}$, the last property means that the linear
transformation \eqref{eq:12} with $D\in\text{SL}(3,\mathbb{C})$ is a Finslerian
isometry of the space $\text{Herm}(3)$, i.e., $|X^\prime|=|X|$. It is clear that
all such isometries form a group. We will give the explicit matrix description
of this group in the basis \eqref{eq:7}.

Let us substitute the expansions $X^\prime=X^{\prime A}\lambda_A$ and $X=X^B\lambda_B$
into \eqref{eq:12}. We then multiply the resulting equality by $\lambda^A$ from the
left, compute its trace, and use the relations \eqref{eq:9}. As a result, we
obtain
\begin{equation}
X^{\prime A}=L(D)^A_B X^B\quad (A,B=0,1,\dots,8),
\label{eq:13}
\end{equation}
where
\begin{equation}
L(D)^A_B=\frac{1}{2}\text{Tr}(\lambda^A D\lambda_B D^+)
\label{eq:14}
\end{equation}
are elements of the matrix of the linear transformation \eqref{eq:12} in the
basis \eqref{eq:7}. It should be noted that $L(D)^A_B\in\mathbb{R}$. Thus, for
any $D\in\text{SL}(3,\mathbb{C})$, the transformation
\eqref{eq:13}--\eqref{eq:14} preserves the form \eqref{eq:11}:
$$
G_{ABC}X^{\prime A}X^{\prime B}X^{\prime C}=G_{ABC}X^A X^B X^C.
$$

Since the group $\text{SL}(2,\mathbb{C})\subset\text{SL}(3,\mathbb{C})$ is
locally isomorphic to the group $\text{O}^\uparrow_+(1,3)$~\cite{Postnikov}, it
is interesting to consider the transformation \eqref{eq:13}--\eqref{eq:14} with
$D\in\text{SL}(2,\mathbb{C})$, i.e., from the point of view of a ``4-dimensional
observer''. This will allow us to represent the expression \eqref{eq:11} for the
Finslerian length of the 9-vector completely in the 4-dimensional form.

Let
\begin{equation}
D_2=
\begin{pmatrix}
d^1_1&d^1_2&0\\
d^2_1&d^2_2&0\\
0&0&1
\end{pmatrix},\quad
\det D_2=1\quad
(d^{\hat a}_{\hat b}\in\mathbb{C}; \hat a,\hat b=1,2).
\label{eq:15}
\end{equation}
The matrices \eqref{eq:15} form a subgroup of $\text{SL}(3,\mathbb{C})$ which is
isomorphic to the group $\text{SL}(2,\mathbb{C})$. Let us substitute the matrix
$D_2$ from \eqref{eq:15} into \eqref{eq:14} instead of $D$. Direct computations
show that
\begin{align}
L(D_2)^0_0&=\frac{1}{2}(
d^1_1\overline{d^1_1}+
d^1_2\overline{d^1_2}+
d^2_1\overline{d^2_1}+
d^2_2\overline{d^2_2}),\notag\\
L(D_2)^0_1&=\frac{1}{2}(
d^1_1\overline{d^1_2}+
d^2_1\overline{d^2_2}+
d^1_2\overline{d^1_1}+
d^2_2\overline{d^2_1}),\notag\\
L(D_2)^0_2&=\frac{i}{2}(
d^1_2\overline{d^1_1}+
d^2_2\overline{d^2_1}-
d^1_1\overline{d^1_2}-
d^2_1\overline{d^2_2}),\notag\\
L(D_2)^0_3&=\frac{1}{2}(
d^1_1\overline{d^1_1}+
d^2_1\overline{d^2_1}-
d^1_2\overline{d^1_2}-
d^2_2\overline{d^2_2}),\notag\\
L(D_2)^1_0&=\frac{1}{2}(
d^1_1\overline{d^2_1}+
d^2_1\overline{d^1_1}+
d^1_2\overline{d^2_2}+
d^2_2\overline{d^1_2}),\notag\\
L(D_2)^1_1&=\frac{1}{2}(
d^1_1\overline{d^2_2}+
d^2_1\overline{d^1_2}+
d^1_2\overline{d^2_1}+
d^2_2\overline{d^1_1}),\notag\\
L(D_2)^1_2&=\frac{i}{2}(
d^1_2\overline{d^2_1}+
d^2_2\overline{d^1_1}-
d^1_1\overline{d^2_2}-
d^2_1\overline{d^1_2}),\notag\\
L(D_2)^1_3&=\frac{1}{2}(
d^1_1\overline{d^2_1}+
d^2_1\overline{d^1_1}-
d^1_2\overline{d^2_2}-
d^2_2\overline{d^1_2}),\notag\\
L(D_2)^2_0&=\frac{i}{2}(
d^1_1\overline{d^2_1}-
d^2_1\overline{d^1_1}+
d^1_2\overline{d^2_2}-
d^2_2\overline{d^1_2}),\notag\\
L(D_2)^2_1&=\frac{i}{2}(
d^1_1\overline{d^2_2}-
d^2_1\overline{d^1_2}+
d^1_2\overline{d^2_1}-
d^2_2\overline{d^1_1}),\notag\\
L(D_2)^2_2&=\frac{1}{2}(
d^1_1\overline{d^2_2}+
d^2_2\overline{d^1_1}-
d^1_2\overline{d^2_1}-
d^2_1\overline{d^1_2}),\notag\\
L(D_2)^2_3&=\frac{i}{2}(
d^1_1\overline{d^2_1}-
d^2_1\overline{d^1_1}-
d^1_2\overline{d^2_2}+
d^2_2\overline{d^1_2}),\notag\\
L(D_2)^3_0&=\frac{1}{2}(
d^1_1\overline{d^1_1}-
d^2_1\overline{d^2_1}+
d^1_2\overline{d^1_2}-
d^2_2\overline{d^2_2}),\notag\\
L(D_2)^3_1&=\frac{1}{2}(
d^1_1\overline{d^1_2}-
d^2_1\overline{d^2_2}+
d^1_2\overline{d^1_1}-
d^2_2\overline{d^2_1}),\notag\\
L(D_2)^3_2&=\frac{i}{2}(
d^1_2\overline{d^1_1}-
d^2_2\overline{d^2_1}-
d^1_1\overline{d^1_2}+
d^2_1\overline{d^2_2}),\notag\\
L(D_2)^3_3&=\frac{1}{2}(
d^1_1\overline{d^1_1}-
d^1_2\overline{d^1_2}-
d^2_1\overline{d^2_1}+
d^2_2\overline{d^2_2}),
\label{eq:16}
\end{align}
$L(D_2)^{3+i}_{3+j}=M(D_2)^i_j$ ($i,j=1,2,3,4$), where
\begin{align}
M(D_2)^1_1&=\frac{1}{2}(\overline{d^1_1}+d^1_1),\quad
M(D_2)^3_1=\frac{1}{2}(\overline{d^2_1}+d^2_1),\notag\\
M(D_2)^1_2&=\frac{i}{2}(\overline{d^1_1}-d^1_1),\quad
M(D_2)^3_2=\frac{i}{2}(\overline{d^2_1}-d^2_1),\notag\\
M(D_2)^1_3&=\frac{1}{2}(\overline{d^1_2}+d^1_2),\quad
M(D_2)^3_3=\frac{1}{2}(\overline{d^2_2}+d^2_2),\notag\\
M(D_2)^1_4&=\frac{i}{2}(\overline{d^1_2}-d^1_2),\quad
M(D_2)^3_4=\frac{i}{2}(\overline{d^2_2}-d^2_2),\notag\\
M(D_2)^2_1&=\frac{i}{2}(d^1_1-\overline{d^1_1}),\quad
M(D_2)^4_1=\frac{i}{2}(d^2_1-\overline{d^2_1}),\notag\\
M(D_2)^2_2&=\frac{1}{2}(d^1_1+\overline{d^1_1}),\quad
M(D_2)^4_2=\frac{1}{2}(d^2_1+\overline{d^2_1}),\notag\\
M(D_2)^2_3&=\frac{i}{2}(d^1_2-\overline{d^1_2}),\quad
M(D_2)^4_3=\frac{i}{2}(d^2_2-\overline{d^2_2}),\notag\\
M(D_2)^2_4&=\frac{1}{2}(d^1_2+\overline{d^1_2}),\quad
M(D_2)^4_4=\frac{1}{2}(d^2_2+\overline{d^2_2}),
\label{eq:17}
\end{align}
$L(D_2)^{8}_{8}=1$, while the other elements of the matrix of the
transformation $X^{\prime A}=L(D_2)^A_B X^B$ vanish. Thus, for
$D=D_2$, the  Finslerian isometry \eqref{eq:13} has the form
\begin{align}
X^{\prime\alpha}&=L(D_2)^\alpha_\beta X^\beta\quad
(\alpha,\beta=0,1,2,3),\notag\\
\theta^{\prime i}&=M(D_2)^i_j\theta^j\quad
(i,j=1,2,3,4),\notag\\
X^{\prime 8}&=X^8,
\label{eq:18}
\end{align}
where $L(D_2)^\alpha_\beta$, $M(D_2)^i_j$ are given by
\eqref{eq:16}--\eqref{eq:17} and the notation $\theta^{\prime i}=
X^{\prime 3+i}$, $\theta^j=X^{3+j}$ is used.

It was shown in the paper \cite{Solo_Vlad:N-spinors} that  \eqref{eq:16} and
\eqref{eq:17} are the elements of the matrices of the transformations for a
Lorentz 4-vector and a Majorana 4-spinor respectively. Therefore, the result
\eqref{eq:18} asserts that, for $D=D_2$, the 9-vector $X^A$ splits into the
Lorentz 4-vector  $X^\alpha$, the Majorana 4-spinor $\theta^i$, and
the Lorentz 4-scalar $X^8$.

This is the essence of the procedure of dimensional reduction allowing to
display the ``4-dimensional structure'' of 9-dimensional expressions. Let us
apply this procedure to the cumbersome formula \eqref{eq:11} for the Finslerian
length of the 9-vector $X^A$. Taking into consideration \eqref{eq:18}, we
obtain
\begin{equation}
|X|^3=g_{\mu\nu}X^\mu X^\nu X^8-g_{\mu\nu}X^\mu\overline{\theta}\gamma^\nu\theta,
\label{eq:19}
\end{equation}
where $\mu,\nu=0,1,2,3$, $\|g_{\mu\nu}\|=\text{diag}\,(1,-1,-1,-1)$ is
the matrix of components of the Minkowski metric tensor in a pseudoorthonormal
basis,
\begin{align*}
&\gamma^0=
\begin{pmatrix}
0&0&i&0\\
0&0&0&-i\\
-i&0&0&0\\
0&i&0&0
\end{pmatrix},\
\gamma^1=
\begin{pmatrix}
i&0&0&0\\
0&-i&0&0\\
0&0&-i&0\\
0&0&0&i
\end{pmatrix},\
\gamma^2=
\begin{pmatrix}
0&i&0&0\\
i&0&0&0\\
0&0&0&i\\
0&0&i&0
\end{pmatrix},\\
&\gamma^3=
\begin{pmatrix}
0&0&-i&0\\
0&0&0&i\\
-i&0&0&0\\
0&i&0&0
\end{pmatrix}\qquad
(\gamma^\mu\gamma^\nu+\gamma^\nu\gamma^\mu=2g^{\mu\nu})
\end{align*}
are the Dirac matrices in the Majorana representation
\cite{Solo_Vlad:N-spinors}, $\theta\in\mathbb{R}^4$ is the 4-component
column of real numbers $\theta^j=X^{3+j}$ ($j=1,2,3,4$), and
$\overline{\theta}\equiv\theta^\top\gamma^0$ (the mark
$\scriptstyle\top$ denotes the matrix transposition). Thus, the
expression \eqref{eq:11} is written in the compact 4-dimensional form
\eqref{eq:19}.
\section*{The generalized Duffin--Kemmer equation}
Let us use the above formalism for the quantum description of a free
3-spinor particle in the 9-dimensional Finslerian space with the
metric \eqref{eq:11}. The corresponding wave equation was obtained in
the works~\cite{Finkelstein:Hypermanifolds, Vlad_Solo:9-Dirac}. The
paper~\cite{Finkelstein:Hypermanifolds} dealt with the coordinate
representation, while the paper~\cite{Vlad_Solo:9-Dirac} did with the
momentum representation of the same wave equation. Below, we use the
momentum representation because the differential equations for wave
functions of free particles become purely algebraic and, therefore,
simpler to analyze in this representation.

Let $i^r$ and $\beta_{\dot s}$ $(r,s=1,2,3)$ be Finslerian 3-spinors,
while $P\equiv\|P^{r\dot s}\|$ is an element of the space
$\text{Herm}(3)$. Substituting the matrix $P$ instead of $X$ into
\eqref{eq:8} and computing the determinant, we obtain
\begin{equation}
\det P=G_{ABC}P^{A}P^{B}P^{C}
\label{(2)}
\end{equation}
in the notation of the formula \eqref{eq:11}. Here, the relations
between $P^{r\dot s}$ and $P^{A}$ have the form
\begin{equation}
\left.\begin{array}{lll}
P^{1\dot 1}=P^0+P^3,&P^{1\dot 2}=P^1-iP^2,&P^{1\dot 3}=P^4-iP^5\\
P^{2\dot 1}=P^1+iP^2,&P^{2\dot 2}=P^0-P^3,&P^{2\dot 3}=P^6-iP^7\\
P^{3\dot 1}=P^4+iP^5,&P^{3\dot 2}=P^6+iP^7,&P^{3\dot 3}=P^8
\end{array}\right\}.
\label{(3)}
\end{equation}

In the work~\cite{Vlad_Solo:9-Dirac}, to describe a free 3-spinor
particle with the wave function $i^r(P^{A})$, $\beta_{\dot s}(P^{A})$,
the $\text{SL}(3,\mathbb{C})$-covariant equation
\begin{equation}
\left.\begin{array}{r@{\;}c@{\;}l}
P^{r\dot s}\beta_{\dot s}&=&Mi^r\\
P_{r\dot s}\,i^r&=&M^2\beta_{\dot s}
\end{array}\right\}
\label{(4)}
\end{equation}
was proposed. Here, $P^{r\dot s}$ are expressed by \eqref{(3)} in
terms of the 9-momentum $P^{A}$ of the particle, $M$ is a positive
scalar, and $P_{r\dot s}$ are the cofactors of the elements
$P^{r\dot s}$ of the matrix $P$. It is natural to call $M$ the 9-mass
of the particle because substituting the upper equality 
of \eqref{(4)} into the lower one (and vice versa) gives a Finslerian
analog of the Klein--Gordon equation for each 3-spinor component of
the wave function:
$$
(G_{ABC}P^{A}P^{B}P^{C}-M^3)i^r=0,\quad
(G_{ABC}P^{A}P^{B}P^{C}-M^3)\beta_{\dot s}=0.
$$

Assuming $P^{3+i}=0$ ($i=1,2,3,4$), $P^8=M$ and performing
dimensional reduction as in \eqref{eq:18}, we conclude that the
equation \eqref{(4)} splits into the standard 4-dimensional Dirac (for
the components $i^1$, $i^2$, $\beta_{\dot 1}$, $\beta_{\dot 2}$) and
Klein--Gordon (for the component $i^3=\beta_{\dot 3}$) equations in
the momentum representation, which describe free particles with the
mass $M$~\cite{Vlad_Solo:9-Dirac}.

It should be noted that the equation \eqref{(4)} is quadratic with
respect to $P^A$. This follows from \eqref{(3)} and the fact that
$P_{r\dot s}$ are proportional to $2\times 2$ minors of the matrix $P$.
Let us try to represent \eqref{(4)} in the form of an equation, which
is linear with respect to the 9-momentum $P^A$.

Following the work~\cite{Solov'yov:Duffin}, we introduce the new
variables $\xi_1$, $\xi_2$, \ldots, $\xi_6$ such that
\begin{equation}
\left.\begin{array}{ll}
P^{2\dot 1}i^1-P^{1\dot 1}i^2=M\xi_1,&P^{2\dot 2}i^1-P^{1\dot 2}i^2=M\xi_4\\
P^{3\dot 1}i^1-P^{1\dot 1}i^3=M\xi_2,&P^{3\dot 2}i^1-P^{1\dot 2}i^3=M\xi_5\\
P^{3\dot 1}i^2-P^{2\dot 1}i^3=M\xi_3,&P^{3\dot 2}i^2-P^{2\dot 2}i^3=M\xi_6
\end{array}\right\}.
\label{(5)}
\end{equation}
With the help of \eqref{(5)}, we can rewrite the lower equality of
\eqref{(4)} in the form
\begin{equation}
\left.\begin{array}{r@{\;}c@{\;}l}
P^{3\dot 3}\xi_4 - P^{2\dot 3}\xi_5 + P^{1\dot 3}\xi_6&=&M\beta_{\dot 1}\\
-P^{3\dot 3}\xi_1 + P^{2\dot 3}\xi_2 - P^{1\dot 3}\xi_3&=&M\beta_{\dot 2}\\
-P^{3\dot 1}\xi_4 + P^{2\dot 1}\xi_5 - P^{1\dot 1}\xi_6&=&M\beta_{\dot 3}
\end{array}\right\}.
\label{(6)}
\end{equation}
Thus, \eqref{(4)} is equivalent to the set of equations $P^{r\dot s}
\beta_{\dot s}=Mi^r$, \eqref{(5)}--\eqref{(6)} or, what is the same,
to the matrix equation
\begin{equation}
\hat P\Psi=M\Psi,
\label{(7)}
\end{equation}
where $\Psi=(i^1,i^2,i^3,\beta_{\dot 1},\beta_{\dot 2},\beta_{\dot 3},
\xi_1,\xi_2,\xi_3,\xi_4,\xi_5,\xi_6)^\top$ is the 12-component column
and
\begin{equation}
\hat P=\left(\begin{array}{cccc}
\bf 0&P&\bf 0&\bf 0\\
\bf 0&\bf 0&P_1&P_2\\
P_3&\bf 0&\bf 0&\bf 0\\
P_4&\bf 0&\bf 0&\bf 0
\end{array}\right)
\label{(8)}
\end{equation}
is the $12\times 12$ matrix consisting of the $3\times 3$ blocks:
\begin{equation}
{\bf 0}=\left(
\begin{array}{ccc}
0&0&0\\
0&0&0\\
0&0&0
\end{array}\right),
\label{(9)}
\end{equation}
\begin{equation}
P=\left(
\begin{array}{ccc}
P^{1\dot 1}&P^{1\dot 2}&P^{1\dot 3}\\
P^{2\dot 1}&P^{2\dot 2}&P^{2\dot 3}\\
P^{3\dot 1}&P^{3\dot 2}&P^{3\dot 3}
\end{array}\right),
\label{(10)}
\end{equation}
\begin{equation}
P_1=\left(
\begin{array}{ccc}
0&0&0\\
-P^{3\dot 3}&P^{2\dot 3}&-P^{1\dot 3}\\
0&0&0
\end{array}\right),
\label{(11)}
\end{equation}
\begin{equation}
P_2=\left(
\begin{array}{ccc}
P^{3\dot 3}&-P^{2\dot 3}&P^{1\dot 3}\\
0&0&0\\
-P^{3\dot 1}&P^{2\dot 1}&-P^{1\dot 1}
\end{array}\right),
\label{(12)}
\end{equation}
\begin{equation}
P_3=\left(
\begin{array}{ccc}
P^{2\dot 1}&-P^{1\dot 1}&0\\
P^{3\dot 1}&0&-P^{1\dot 1}\\
0&P^{3\dot 1}&-P^{2\dot 1}
\end{array}\right),
\label{(13)}
\end{equation}
\begin{equation}
P_4=\left(
\begin{array}{ccc}
P^{2\dot 2}&-P^{1\dot 2}&0\\
P^{3\dot 2}&0&-P^{1\dot 2}\\
0&P^{3\dot 2}&-P^{2\dot 2}
\end{array}\right).
\label{(14)}
\end{equation}

Let us raise \eqref{(8)} to the fourth power. The direct calculation
shows that
\begin{equation}
{\hat P}^4=(\det P)\hat P.
\label{(15)}
\end{equation}
On the other hand, by using \eqref{(3)} and \eqref{(9)}--\eqref{(14)},
it is easy to represent \eqref{(8)} in the form of the linear
combination
\begin{equation}
\hat P=P^{A}\delta_{A}
\label{(16)}
\end{equation}
of the nine $12\times12$ matrices $\delta_{A}$ (in the explicit form,
these matrices are given in the Appendix to the
paper~\cite{Solov'yov:Duffin}). The substitution of \eqref{(2)} and
\eqref{(16)} into \eqref{(15)} results in the identity
\begin{equation}
(P^{A}\delta_{A})^4=G_{ABC}P^{A}P^{B}P^{C}(P^{D}\delta_{D})
\label{(17)}
\end{equation}
valid for any $P^{A}$. Here, $A,B,C,D=0,1,\dots,8$.

It is evident that \eqref{(17)} generalizes the known 4-dimensional
identity
$$
(p^\mu\beta_\mu)^3=g_{\mu\nu}p^\mu p^\nu(p^\lambda\beta_\lambda),
$$
where $\mu,\nu,\lambda=0,1,2,3$ and $\beta_\mu$ are the Duffin--Kemmer
matrices~\cite{Duffin, Kemmer}. Moreover, it follows from \eqref{(17)}
that the $\delta$-matrices satisfy the conditions
\begin{equation}
\delta_{(A}\delta_{B}\delta_{C}\delta_{D)}=
6\{G_{ABC}\delta_{D}+G_{ABD}\delta_{C}+
G_{ACD}\delta_{B}+G_{BCD}\delta_{A}\},
\label{(18)}
\end{equation}
where the parentheses denote the symmetrization with respect to all the
subscripts (i.e., the sum over all permutations of $A,B,C,D$). In this
connection, it is interesting to recall important relations of the
Duffin--Kemmer algebra:
\begin{equation}
\beta_{{(}\mu}\beta_\nu\beta_{\lambda{)}}=2\{g_{\mu\nu}\beta_\lambda +
g_{\lambda\mu}\beta_\nu + g_{\lambda\nu}\beta_\mu\}.
\label{(19)}
\end{equation}
It is easy to see the full analogy between the formulas \eqref{(18)}
and \eqref{(19)}.

Let us return to the equation \eqref{(7)}. With the help of
\eqref{(16)}, it can finally be written in the following form
\begin{equation}
(P^{A}\delta_{A} - M)\Psi=0,
\label{(20)}
\end{equation}
where $\delta_{A}$ satisfy the conditions \eqref{(18)}. Thus, the purpose of
this section is achieved: the equation \eqref{(4)} is represented in
the form of the  generalized Duffin--Kemmer equation \eqref{(20)}.
\section*{Conclusion}
Summarizing, we make some remarks concerning the obtained results.

In this paper, the main facts of the geometry of Finslerian 3-spinors
of the 9-dimensional linear space with the metric function defined by
the cubic form \eqref{eq:11} are formulated. The explicit description
of isometries of this 9-dimensional Finslerian space and the procedure
of dimensional reduction, which allowed us to represent \eqref{eq:11}
in the 4-dimensional form \eqref{eq:19}, are given. The latter is
important because it demonstrates the correspondence of our
constructions to the standard relativistic theory on the level of
geometry.

In addition, we deduced the generalized Duffin--Kemmer equation
\eqref{(20)} for a free Finslerian 3-spinor particle in the momentum
representation. In a parallel way, we obtained the 9-dimensional
Finslerian analog \eqref{(18)} of the defining relations \eqref{(19)}
of the 4-dimensional Duffin--Kemmer algebra. It is also shown that the
equation \eqref{(20)} unifies the 4-dimensional  Dirac and Klein--Gordon
equations in a nontrivial way.

The author is grateful to Professor Yu.~S.~Vladimirov for the fruitful
collaboration during many years, which is expressed in the number of
the common papers.

\end{document}